\documentclass[twocolumn,english,showpacs,floatfix,amsmath,amssymb,prb,aps,superscriptaddress] {revtex4}
\usepackage[T1]{fontenc}
\usepackage[latin9]{inputenc}
\usepackage{babel}

\usepackage{graphicx}
\usepackage{amssymb}
\usepackage{amsmath}
\usepackage{esint}

\newcommand{\be}{\begin{equation}}
\newcommand{\ee}{\end{equation}}
\newcommand{\ba}{\begin{eqnarray}}
\newcommand{\ea}{\end{eqnarray}}

\begin{document}

\title{Chiral heat transport in driven quantum Hall and quantum spin Hall edge states }
\author{ Liliana Arrachea}

\affiliation{Departamento de F\'{\i}sica, Facultad de Ciencias Exactas y Naturales, Universidad de Buenos Aires,
Pabell\'on 1, Ciudad Universitaria, 1428, Buenos Aires, Argentina.}

\author{ Eduardo Fradkin}

\affiliation{Department of Physics, University of Illinois at Urbana-Champaign, 1110 West Green Street, Urbana, Illinois 61801-3080, USA.}

\begin{abstract}
We consider 
a model for an edge state of electronic systems in the quantum Hall regime with
filling $\nu=1$ and in the quantum spin Hall regime. In both cases the system is in contact with two reservoirs by tunneling at point contacts. Both systems are locally driven
by applying an ac voltage in one of the contacts. By weakly coupling them to a third reservoir, the 
transport of the generated heat is  studied in two different ways: i) when the third reservoir acts as a thermometer
the local temperature is sensed, and ii) when the third reservoir acts as a voltage probe the time-dependent local voltage is sensed. Our results  indicate a  chiral propagation of the heat along the edge in the quantum Hall case and in the quantum spin Hall case (if the injected electrons are spin polarized). We also show
that a similar picture is obtained if instead of heating by ac driving the system is put in contact  
to a stationary reservoir at a higher temperature. In both cases the temperature profile shows that the electrons along the
edge thermalize with the closest  ``upstream''  reservoir.
\end{abstract}

\maketitle

\section{Introduction}

A key and almost defining property of quantum Hall states, integer\cite{Halperin-1982} and fractional,\cite{Wen-1990}  is the existence of a structure of 
gapless states at the boundary of these incompressible quantum fluids which propagate only along a direction dictated by the 
external perpendicular magnetic field, chiral edge states. 
 The experimental detection of the chiral nature of these states is thus crucial to the understanding of these fluids and
 has been and is the focus of intense research, in experiment and theory alike.\cite{QH-edges} The recently discovered of  two-dimensional topological 
 insulators, 
 strong spin-orbit coupled semiconductors that exhibit a quantized anomalous Hall effect (if spin polarized) or the quantum spin Hall effect 
 (if unpolarized), are also 
 predicted to have a universal edge structure.\cite{QSH}  In the anomalous quantum Hall state the edge states are chiral whereas in the quantum spin 
 Hall effect the 
 edge states with opposite polarization propagate in opposite directions.

In a recent very interesting experiment,  evidence of chiral propagation of the heat 
along an edge state  in a GaAs/AlAs heterostructure
 with a two-dimensional electron gas in the integer quantum Hall regime  has been presented. \cite{granger} The experiment was performed in
the quantum Hall regime with filling $\nu=1$ locally heated by an ac field. Heat transport 
in ac driven systems have been recently the focus of experimental and theoretical interest  in 
several electronic, \cite{pumheat,lilileto,tempe1} 
phononic \cite{heatphon} and photonic \cite{heatphot} systems. 
So far experiments of this type have not been done in graphene-based devices, which have the advantage that the integer quantum Hall effect is seen  
at room 
temperature \cite{QH-graphene},  or in two-dimensional topological insulators.
The goal of the present work is the theoretical analysis of a setup close to the experimental
work for the integer quantum Hall case.\cite{granger} Our model and results should also apply to graphene, and we also generalize it to quantum spin 
Hall systems. 
As we will see below, our results verify the empirical conclusions of Ref. [\onlinecite{granger}]  regarding 
the chiral propagation of the heat along the edge. Another conclusion
of that work is the electronic cooling in the propagation along the edge. Our results 
indicate that the propagation is coherent along the edge, the electrons preserving some ``memory''
of the temperature of the last reservoir they have visited.

\begin{figure}[tb]
\includegraphics[width=.48\textwidth]{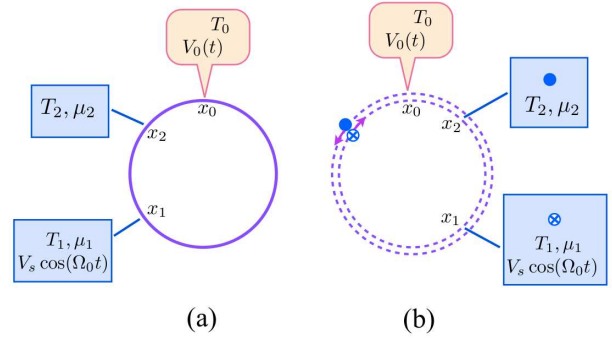}
 \caption{Sketch of the considered setup. (a) The quantum Hall edge state is represented
by a circle. Two reservoirs are connected at the positions $x_1$ (source) and $x_2$
(drain). An ac voltage is applied at the source reservoir. A third reservoir is weakly connected at $x_0$ in order to sense the local voltage $V_0(t)$ or the local temperature $T_0$. (b) Quantum spin Hall effect. The system contains a pair of edge states where electrons with spin up (down) move clockwise (counterclockwise). The source and drain reservoirs are spin-polarized. }
\label{fig1}
\end{figure}

We base our study in a microscopic model for the edge, consistent in a ring of free chiral fermions
 connected to two fermionic reservoirs through tunneling couplings, as indicated in
the sketch of Fig.\ref{fig1}. In the case of the spin Hall case, sketched in Fig.1b, we consider two states in the edge, corresponding
to electrons moving with opposite chirality and helicity. We also consider the possibility of spin-polarized reservoirs in the latter case.
 In one of the reservoirs, an ac voltage is applied, acting as a local heater.
We consider a weak coupling to a third reservoir which will be used to define a voltage probe or a thermometer.
Both kinds of probes  are used to sense the local temperature along the edge. In the case of the voltage probe,
the signal of the time-dependent voltage is experimentally used to get an estimate of the local temperature from thermoelectric 
effects. \cite{granger,molen} A thermometer is defined in a gedanken setup, where the temperature of a weakly coupled
reservoir is fixed form the condition of  a vanishing heat flow through the contact to the system under investigation. 
That definition of temperature was
originally proposed  in Ref. \onlinecite{engan} in the context of stationary electronic transport. It as been
adopted to analyze stationary transport \cite{dubi-diventra} in nano-devices and generalized to the context of 
 systems under AC driving. \cite{tempe1} This definition of the local temperature is correct is the system is weakly driven,
and  agrees with other definitions of the temperature from fluctuation-dissipation relations in non-equilibrium
systems. \cite{tempe1,lilileto,fdr}

This work is organized as follows. In section \ref{sec:model} we present a simple model that mimics the setup of Granger {\it et al}.\cite{granger} and we
generalize it for the case of spin-Hall effect.
In section \ref{sec:currents} we derive expressions for the heat current in  terms of Green functions of chiral fermions in this non-equilibrium situation. 
In sections \ref{sec:voltage} and \ref{sec:thermometer} we present, respectively, a theory of sensing with voltage probes and thermometers along the edge. 
In section \ref{sec:res} we present results for the heat propagation in the cases of the quantum Hall as well as in the quantum spin Hall effect.
 Section \ref{sec:conclusions} is devoted to the conclusions.

\section{Model}
\label{sec:model}

The full system is described by the following Hamiltonian:
\be \label{ham}
H=H_{edge}+\sum_{j=0}^2[H_{j}+H_{c,j}] + H_{AC}(t),
\ee
where the edge is represented by a ring of circumference $L$ along which
chiral fermions circulate with velocity $v_F$. In the case of topological insulators (the spin quantum Hall state), there are two 
edge states (Kramers pairs) with opposite chiralities and helicities (which we will refer to as ``spin''). The
Hamiltonian for the edge states in the latter case is
\ba
H_{edge} &= & \sum_{\sigma  =  \pm }\int_{0}^L dx \Psi_{\sigma}^{\dagger}(x) {\cal D}_x
\partial_x \Psi_{\sigma}(x)\nonumber \\
& = & \sum_{p,\sigma} v_{F,\sigma} (p-\frac{\Phi}{L}) \Psi^{\dagger}_p \Psi_p,
\ea
where ${\cal D}_x= - i  v_{F, \sigma} \partial_x- \Phi/L$, with $\Phi$ the magnetic flux threading the ring
in units of the flux quantum $e h/c$. We assume that the  electrons moving clockwise have spin $\uparrow$, while those 
moving counterclockwise have spin $\downarrow$. Thus,
 $v_{F,\sigma}= \sigma v_F$, with $p= 2 n \pi/L$, where $n$ is an integer and $|p|\leq K$,
while 
\be
H_j=-i \sum_{\sigma= \pm} v_{F,\sigma}^{j}
\int_0^{\infty} d r_{j}
\Psi^{\dagger}_{\sigma}(r_{j}) \partial_{r_{j}} \Psi_{\sigma}(r_{j}),
\ee
are the Hamiltonians of infinite systems of chiral fermions, which
play the role of reservoirs. The source ($j=1$) and drain ($j=2$) reservoirs
are at  temperatures $T_1$ and $T_2$, respectively and have the same chemical potential $\mu$. 
We consider the possibility for the reservoirs $j=1,2$ to being spin-polarized, which is equivalent to assuming 
$v_{F,+}^{j} \neq v_{F,-}^{j}, \;\;j=1,2$.
The system is driven by applying an
AC voltage $V_1(t)= V_S \cos(\Omega_0 t )$ at the reservoir $j=1$. The reservoir  
$j=0$ corresponds to a probe that may act as a thermometer or a voltage probe. 
In the first case
it has the
same chemical potential $\mu$ as the other reservoirs, while its temperature
$T_0$ is adjusted in order to satisfy the condition of vanishing heat flow
along its contact to the ring. In the second case, it has the same temperature as the other reservoirs,
while it has an AC voltage $V_0(t)= \mu_0 + \sum_{k \neq 0}  e^{-i (k \Omega_0 t+ \varphi_k) } V_0^{(k)}/2 $. 
The different harmonics $V_0^{(k)}$ are adjusted to satisfy the condition of a vanishing charge flow
along the contact.

The time-dependent voltages can be described in terms of the Hamiltonian
\be
H_{AC}(t)= \sum_{j=0,1,\sigma} V_j(t) \int \frac{dp_j}{2 \pi}  \Psi^{\dagger}_{p_j,\sigma} \Psi_{p_j,\sigma}
\ee

The contacts are described by the Hamiltonians:
\be \label{cont}
H_{c,j} = w_{j} \sum_{\sigma} [\Psi^{\dagger}_{\sigma}(x_{j}) \Psi_{\sigma}(r^0_{j})+ H. c.],
\ee
where $x_{j}$ and $r^0_{j}$ are, respectively the positions of the ring 
and the reservoir
at which the contact is established. We assume that the tunneling parameter
$w_0$ between the ring and the probe reservoir is so weak that it  introduces
negligible dephasing in the particle propagation along the ring.  

In the case of the usual quantum Hall effect, there is just one edge state and the spin label becomes
irrelevant. Thus, in order to model that case, we should consider the above Hamiltonian with just one of the helicities $\sigma$. 

In the following sections we will compute the local temperature by sensing with a thermometer, and voltage by sensing with a voltage probe, along the edges for each 
of these systems.

\section{Charge and heat currents through the contact to the probe} 
\label{sec:currents}

The expression for the charge and heat currents flowing through the contact between
the edge and the probe reservoir are obtained from the general laws of
the conservation of the charge and energy, respectively,
\begin{widetext}
\ba \label{jc}
\frac{ d \langle N_0 \rangle}{dt} & = &  J^c_0(t)=2 w_0 \sum_{p_0,\sigma}  \mbox{Re} \{ e^{-i p_0 r_0^0} G_{x_0,p_0,\sigma}^<(t,t) \},\nonumber \\
\frac{ d \langle H_0 -\mu N_0 \rangle}{dt} & = &  J^Q_0(t) =  2 w_0 \sum_{p_0,\sigma}
 \mbox{Re} \{ e^{-i p_0 r_0^0} (\varepsilon_{p_0,\sigma} - \mu) G_{x_0,p_0,\sigma}^<(t,t) \},\nonumber
\ea
\end{widetext}
where $N_0 = \sum_{p_0,\sigma} \Psi^{\dagger}_{p_0,\sigma} \Psi_{p_0,\sigma}$, $\varepsilon_{p_0,\sigma}= \sigma v_F^0 p_0$, while we have introduced
the lesser Green function $G^<_{x_0,p_0,\sigma}(t,t^{\prime})= 
i \langle   \Psi^{\dagger}_{p_0,\sigma}(t^{\prime}) \Psi_{\sigma}(x_0,t) \rangle $. 
Since both currents are generated by an AC voltage of frequency $\Omega_0$ they can be in general expressed as
\begin{equation}
 J^{c,Q}_0(t) = \sum_k e^{-i k \Omega_0 t} J^{c,Q}_0(k).
 \end{equation}

The lesser Green function satisfies the following Dyson equation
\ba \label{dymix}
G^<_{x_0,p_0,\sigma}(t,t^{\prime}) & = & w_0 \int_{-\infty}^{+\infty} dt_1
[ G^<_{x_0,x_0,\sigma} (t,t_1) g^A_{k_0,\sigma}(t_1,t^{\prime})  \nonumber \\
& + & G^R_{x_0,x_0,\sigma} (t,t_1) g^<_{p_0,\sigma}(t_1,t^{\prime}) ],
\ea
where we have introduced the retarded Green function $G^R_{x,x^{\prime},\sigma}(t,t^{\prime}) = - i \Theta(t-t^{\prime}) 
\langle  \{ \Psi_{\sigma}(x,t) ,  \Psi^{\dagger}_{\sigma}(x^{\prime},t^{\prime}) \}  \rangle$, as well as 
\begin{equation}
 g^{R  (<)}_{p_0,\sigma}(t,t^{\prime})= g^{0,R (0,<)}_{p_0,\sigma}(t-t^{\prime}) \phi_0(t,t^{\prime}),
\end{equation}
where 
\begin{align}
g^A_{p_0,\sigma}(t,t^{\prime})&= [g^R_{p_0,\sigma}(t^{\prime},t)]^*
\nonumber\\
g^{0,R}_{p_0,\sigma}(t-t^{\prime})  &= -i \Theta(t-t^{\prime}) \exp{\{-i \varepsilon_{p_0,\sigma}(t-t^{\prime})\}}
\nonumber\\
g^{0,<}_{p_0,\sigma}(t-t^{\prime})  &= i 2 \pi f(\varepsilon_{p_0,\sigma}-\mu_0) \exp \{-i \varepsilon_{p_0,\sigma}(t-t^{\prime}) \}.
\end{align}
Here $f(\varepsilon_{p_0,\sigma}-\mu_0)$ is the Fermi function with a chemical potential $\mu_0$ and temperature $T_0$. 
The function 
$\phi_0(t,t^{\prime}) $ contains information on the AC-potentials applied at the probe
\begin{eqnarray}
\phi_0(t,t^{\prime}) & = & 
 \exp \{-i \sum_{k=1}^2 V_0^{(k)} \int_{t^{\prime}}^t dt_1 \cos(k \Omega_0 t_1 + \varphi_k) \} \nonumber \\
 & &
\simeq 1-i \sum_{k=1}^2 V_0^{(k)} \int_{t^{\prime}}^t dt_1 \cos(k \Omega_0 t_1 + \varphi_k)+\ldots,\nonumber\\
&&
\end{eqnarray}
where in the second line we have assumed that the amplitudes $V_0^{(k)}$ are low enough. For the probe acting as a thermometer $V_0^{(k)}=0$ and $\mu_0=\mu_1=\mu_2 \equiv \mu$. 
 Thus, $\phi_0(t,t^{\prime})=1$ while $T_0$ defines
the sensed local temperature, which is determined from the condition $J^{Q}_0(0)=0$.\cite{tempe1}
For the probe acting as a voltage probe, we consider $T_0=T_1=T_2 \equiv T$, while
$\mu_0, V_0^{(k)}, \varphi_k$ are determined by demanding the conditions $J^c_0(k)=0, \;\; k=-2, \ldots,2$. \cite{fourpointfed}

Following the procedure of Ref. \onlinecite{liliflo}, it is convenient to take explicitly into account the harmonic time-dependence of the
problem, considering the following representation for the 
Green functions:
\begin{equation} \label{flo}
G^{R,<}_{x,x^{\prime},\sigma}(t,t^{\prime})  =  \sum_k e^{- i k \Omega_0 t} \int \frac{d \omega}{2 \pi}
{\cal G}^{R,<}_{x,x^{\prime},\sigma}(k,\omega) e^{- i \omega (t- t^{\prime})}.
\end{equation}

In the evaluation of the Green functions for coordinates within the ring,  it is also convenient to integrate out
the degrees of freedom of the reservoirs by defining the following ``self-energies'':
\begin{equation}
\Sigma^{R,<}_{j,\sigma} (t,t^{\prime})   =   i 
\phi_j (t, t^{\prime}) \int \frac{d \omega}{ 2 \pi} \lambda_j^{R,<}(t- t^{\prime}, \omega)
\Gamma_{j,\sigma} (\omega) e^{-i \omega (t- t^{\prime})}, 
\end{equation}
where  $\lambda^{R}_j(t- t^{\prime}, \omega) =-i
\Theta(t- t^{\prime})$,  $\lambda^{<}_j(t- t^{\prime},\omega) =i f_j (\omega)
\Theta(t- t^{\prime})$,  with
$f_j (\omega)$  the Fermi function, which depends on the temperature and the chemical
potential of the reservoir while the functions $ \phi_j  (t, t^{\prime}) $ take into account the
applied ac voltages. For our configuration $\phi_2(t, t^{\prime})=1$, while 
\begin{eqnarray}
\phi_1(t,t^{\prime}) &= & \exp \{-i V_S \int_{t^{\prime}}^t dt_1 \cos(\Omega_0 t_1) \}
\sim 1 -   i V_S \times 
\nonumber \\
&  & \int_{t^{\prime}}^t dt_1 \cos(\Omega_0 t_1) - \frac{V_S^2}{2}
\left[ \int_{t^{\prime}}^t dt_1 \cos(\Omega_0 t_1) \right]^2,
\nonumber\\
&&
\end{eqnarray}
where we have assumed in the second step that the amplitude $V_S$ is small.

\section{Sensing with a voltage probe}
\label{sec:voltage}

Substituting Eq. \eqref{dymix} and the representation Eq. \eqref{flo} in the expression of Eq.\eqref{jc} for the charge current results 
\begin{widetext}
\begin{eqnarray}\label{jp}
& & J^c_0(t)  =2 \mbox{Re} \Big\{\sum_{k,\sigma} e^{-i k \Omega_0 t}
\int \frac{d \omega}{2 \pi} \Big[ {\cal G}^<_{x_0,x_0,\sigma}(k,\omega) \Sigma_{0}^A(\omega) + 
 {\cal G}_{x_0,x_0,\sigma}(k,\omega) \Sigma_{0}^<(\omega) \Big] 
 +  \sum_{ s = \pm 1} \sum_{n=1}^2  s \frac{V_0^{(n)}}{2\Omega_0} e^{-i s \varphi_n} \times \nonumber \\
 & & 
 \left\{
\Big[  {\cal G}^<_{x_0,x_0,\sigma}(k- s n,\omega_{s n} )- {\cal G}^<_{x_0,x_0,\sigma}(k- s n,\omega)\Big]  \Sigma_{0}^A(\omega) +
 \Big[{\cal G}_{x_0,x_0,\sigma}(k- s n,\omega_{s n } ) 
 - {\cal G}_{x_0,x_0,\sigma}(k- s n,\omega)\Big]  \Sigma_{0}^<(\omega)  \right\}  \Big\},
\end{eqnarray}
\end{widetext}
where 
\begin{align}
\Sigma_0^A(\omega&) = w_0^2 \sum_{k_0} [g_{k_0}^0(\omega)]^* \sim i \Gamma_0/2
\nonumber\\
\Sigma_0^<(\omega) &= i f_0(\omega)  w_0^2 \sum_{k_0} \delta(\omega-\varepsilon_{k_0})\sim
 i f_0(\omega) \Gamma_0
 \end{align}
  with $f_0(\omega)= f(\omega-V_0)$ and $\Gamma_{0,\pm}=\Gamma_0$.

Since we consider a weak coupling $w_0$, we shall neglect terms $\propto w_0 $ in the evaluation of the functions
${\cal G}_{x_0,x_0,\sigma} (k,\omega) $ and ${\cal G}^<_{x_0,x_0,\sigma} (k,\omega) $.
 This corresponds to evaluating these Green functions 
considering just the
coupling to the source reservoir and neglecting the coupling to the probe and the resulting current $ J^c_0(t)$ is exact up to ${\cal O}(w_0^2)$. 
For the source reservoir, we can 
define $\Sigma_{j,\sigma}^A(\omega) \sim i \Gamma_{j,\sigma} /2, \;\;j=1,2$,
and 
$\Sigma_{j,\sigma}^<(\omega) = i f_{j}(\omega)   \Gamma_{j,\sigma}$, 
where $f_j(\omega)= 1/(1+e^{\beta_j(\omega - \mu)})$, with $\beta_j=T_j^{-1},\;\;j=1,2$.

After some algebra, for low driving frequency $\Omega_0$, the solution of the set of conditions $J_0^c(k)=0,\;\;k=-2, \ldots, 2$ yield the results for the chemical potential and voltage profiles (to order ${\cal O}(V_S^2)$)
\begin{eqnarray} \label{volt}
 \mu_0  & = & \mu-(\frac{V_S}{2})^2 \frac{ \alpha^{\prime}(\mu)}{\rho_{x_0}(\mu)},\nonumber \\
 V_0^{(1)} & = & V_S \frac{\alpha(\mu)}{\rho_{x_0}(\mu)}, \nonumber \\
 V_0^{(2)} & = & \frac{1}{2} (\frac{V_S}{2})^2 \frac{\alpha^{\prime}(\mu)}{\rho_{x_0}(\mu)}, 
\end{eqnarray}
where $\rho_{x_0}(\omega)= -2 \sum_{\sigma} \mbox{Im}[G^0_{x_0,x_0,\sigma}(\omega)]$ and $\alpha(\omega)=\sum_{\sigma} |G^0_{x_0,x_1,\sigma}(\omega)|^2  \Gamma  $,
where $G^0_{x,x^{\prime},\sigma}(\omega)$ is the equilibrium retarded Green function of the edge connected to the reservoirs
$j=1,2$. 
Notice that the behavior of these three harmonics is not independent one another, since $\mu_0-\mu = - 2 V_0^{(2)} $, where 
these quantities are determined by the value of the  function $\alpha(\omega)$ evaluated at $\mu$, while 
$V_0^{(1)}$ is determined by the derivative of this function.

\section{Sensing with a thermometer}
\label{sec:thermometer}

The dc component  of the heat current defined in section \ref{sec:currents} can be written as follows
\begin{eqnarray}
J^Q_0(0) & = &   \Gamma_0 \int \frac{d \omega}{2 \pi}
(\omega - \mu)
\{
i {\cal G}^<_{x_0,x_0,\sigma}(0,\omega) - \nonumber \\
& &  2 f_0(\omega) \mbox{Im} [{\cal G}_{x_0,x_0,\sigma}(0,\omega)] \}
\end{eqnarray}
Calculating the Green functions ${\cal G}^<_{x_0,x_0,\sigma}(0,\omega)$, while
 keeping terms up to ${\cal O}(V_S^2)$ and expanding in $\Omega_0$ and $T_0$,
the condition $J^{Q}(0)=0$ now lead to the result
\begin{equation} \label{t0}
T_0^2=  \frac{ [ T_1^2+ 3 V_S^2/(2 \pi^2) ] \alpha_1^{\prime}(\mu)  +  T_2^2 \alpha_2^{\prime}(\mu) }{\alpha_0^{\prime}(\mu)},
 \end{equation} 
where $T_j$ are the temperatures of the reservoirs $j=1,2$, respectively, while
\begin{eqnarray}
\alpha_0 (\omega) & = &  (\omega-\mu)\rho_{x_0}(\omega) ,\nonumber \\
\alpha_j (\omega) & = &(\omega-\mu) \sum_{\sigma}  |G^0_{x_0,x_j,\sigma}(\omega)|^2 \Gamma_{j,\sigma}, \;\;j=1,2.
\end{eqnarray}

It is interesting to mention that in the limit where $V_S \rightarrow 0$, the temperature $T_0$ defined in
Eq. \eqref{t0} reduces to the local temperature sensed by a thermometer when the heat transport is induced in purely stationary conditions
 by connecting the ring to
reservoirs with different temperatures $T_1 \neq T_2$. Thus,
the ac driving renormalizes the temperature of the reservoir at which it is applied by a factor $\propto V_S$. In particular,
for reservoirs at equal temperature $T_1=T_2=0$, the effect of the ac driving is equivalent to having the source reservoir 
at a temperature $T_1= \sqrt{3/2} V_S/\pi$.

Another interesting feature is the fact that we can approximate the functions $\alpha_0^{\prime}(\mu) \approx \rho_{x_0}(\mu)$
and  $\alpha_1^{\prime}(\mu) \approx \alpha (\mu)$. For a source reservoir at $T_2=0$, we then find:
\begin{equation}
T_0=\frac{T_1}{\sqrt{\rho_{x_0}}} +\frac{3 V_S \sqrt{\rho_{x_0}}  }{4 \pi^2} V_0^{(1)},
\end{equation}
which suggest a rather straightforward relation between the local temperature and the first harmonic of the local voltage.
This is somehow in contrast with the assumption done in the experimental work, that the second harmonic is more sensitive
to the heat transport that the first one. We will show in the next section, however, that the three harmonics analyzed in the
present work contain the relevant signatures for the behavior of the heat propagation along the edge.

\section{Results} \label{sec:res}
\subsection{Quantum Hall edge}
We begin analyzing the usual quantum Hall case, which corresponds to the Hamiltonian of Eq.\eqref{ham} with a single chirality, which
we assume to be $\sigma=+$.
The unperturbed retarded Green function $G^0_{x,x^{\prime},\sigma}(\omega)$ is evaluated in Appendix \ref{apa}. The expression for 
 of the harmonics of the sensed local voltage, Eq.\eqref{volt}, involve the local density of state $\rho_{x_0}(\omega)= \Theta(\Lambda -|\omega|)\pi/\Lambda$ 
 (where $\Lambda$ is a high-energy cutoff), and the function $\alpha(\omega)$ and $\gamma(\omega)$, which in this case are given by 
\[ \alpha(\omega)= \left\{ \begin{array}{ll}
        \gamma(\omega) & \mbox{if $x<x_1, \;\; x \geq x_2$};\\
\gamma(\omega)      
(1+\overline{\Gamma}_{2,+}) & \mbox{if $x_1 \leq x < x_2 $}.\end{array} \right. \] 
and  
\begin{equation}
\gamma(\omega)= \frac{\pi \overline{\Gamma}_{1,+}}{4 \Lambda |\Delta_{+}(\omega)|^2 \sin^2(\frac{\omega L}{2 v_F}+\frac{\Phi}{2})}.
\end{equation}

\begin{figure}[t!]
 \centering
 \includegraphics[width=0.45\textwidth]{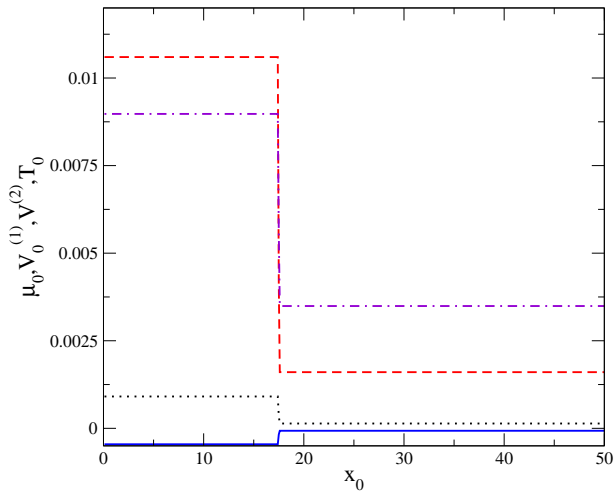}
 \caption{Sensed harmonics of the local time-dependent potential $ \mu_0-\mu$ (dotted line), $V_0^{(1)}$ (dashed line) and $V_0^{(2)}$ (solid line), and local temperature $T_0$ (dashed-dotted line) 
 as functions of the position of the voltage probe
$x_0$ along a ring of $L=50$ with electrons moving clockwise with Fermi velocity $v_F=1$. The source and drain reservoirs are connected
at $x_1=0$ and $x_2=0.35L$, respectively. The chemical potential is $\mu=0.05$, the coupling parameters are
$\overline{\Gamma}_{1,+}=\overline{\Gamma}_{2,+}=\pi$, the temperatures of the reservoirs are $T_1=T_2=0$ and the  amplitude of the ac voltage applied 
at the source is $V_S= 0.05$.}
\label{fig2}
\end{figure}

Thus, for fixed values for the length  of the edge $L$, the Fermi velocity of the electrons along the edge, $v_F$ and the chemical potential $\mu$,
the different harmonics of the local voltage  have piecewise constant profiles as functions of the position
of the voltage probe, the amplitude being  a factor $(1+\overline{\Gamma}_{2,+}) $ larger upstream than downstream, while they display discontinuities
at the positions $x_1$ and $x_2$ at which the source and drain reservoirs are coupled.
The same behavior is found for the local temperature (\ref{t0}) sensed by the thermometer. In fact, assuming $T_2=0$, 
$\alpha_1^{\prime}(\omega) \sim \alpha(\omega)$
and $\alpha_0^{\prime}(\omega)=\rho_{x_0}(\omega)$. Thus, the local temperature is a factor $\sqrt{1+\overline{\Gamma}_{2,+}}$ higher upstream than downstream.
Typical profiles are shown in Fig. \ref{fig2}. In the case of considering a finite cutoff in the energy spectrum of the edge, the three harmonics of the local voltage, as well as the local temperature display oscillations as functions of $x_0$, which are mounted in the stepwise profile. The frequency and amplitude of these oscillations go to zero
in the limit of infinite cutoff considered in   Fig. \ref{fig2}. 
The behavior observed  in this Fig. is exactly the opposite when the movements of the
electrons is inverted. This behavior is consistent with electrons heated 
by the ac voltage and the ensuing current injected through the contact to
the source reservoir. The heated electrons propagate chirally until they
reach the drain electrode, where they tend to thermalize to the temperature of
this reservoir, by means of inelastic scattering processes, with particles and
energy exchanged at the corresponding contact. The net flow into the edge
keeps propagating chirally at an effective temperature that is close to the one of
of the drain reservoir until they reach again the source reservoir. 

\begin{figure}[hbt]
 \centering
 \includegraphics[width=0.4\textwidth]{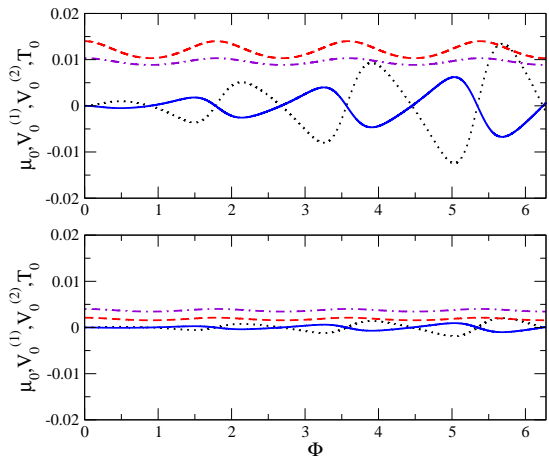}
 \caption{Sensed dc component of the local voltage $\mu_0-\mu$ (dotted line), first harmonic $V_0^{(1)}$ (dashed line),
second harmonic $V_0^{(2)}$ (solid line) and local temperature $T_0$ (dashed-dotted line) as functions of the applied magnetic flux $\Phi$
for ${\cal E}=1$.
 The remaining parameters are the same of Fig.
\ref{fig2}. The upper panel corresponds to sensing upstream
($x_1<x_0<x_2$), while the lower panel corresponds to sensing downstream ($x_0>x_2$).
}
\label{fig3}
\end{figure}

We also notice that the local voltage and temperature display a non-trivial behavior as functions of the magnetic field $B$. The magnetic field
  enters in the function $|\Delta_{+}(\omega)|^2 \sin^2(\frac{\omega L}{2 v_F}+\frac{\Phi}{2})$ of the denominator
of $\gamma(\omega)$, through the magnetic flux $\Phi$ as well in the field-dependence  of the Fermi velocity, $v_F= {\cal E}/\Phi$,
with ${\cal E}= E L^2 c/(4 \pi e h)$, being $E$ the electric field. 
In Fig. \ref{fig3} we show the  different harmonics of the time-dependent local voltage, as well as the local temperature
as functions of $B$. We consider fixed
$\mu$, $L$ and positions $x_1$ and $x_2$ for the source and drain connections. The upper panel corresponds to sensing at a position $x_1 <x_0 < x_2$ 
(upstream) while the lower panel corresponds to a downstream position of the probes. 
The behavior of the harmonics $V_0^{(1)}$ and $V_0^{(2)}$ of the local voltage qualitatively resembles that observed in Fig. 3 of the experimental work [\onlinecite{granger}].
Our results show that the ``amplification factor'' in the upstream signals relative to the downstream ones is fully determined by the degree of coupling of the drain reservoir, represented by $\overline{\Gamma}_{2,+}$.
As stressed in the previous section, such amplification, which is the signature of the chiral propagation of the charge and
heat currents along the edge, are observed in all the harmonics of the local voltage and the local temperature alike. 
In fact, the amplification is observed in  the experimental work  in two harmonics $V_0^{(1)}$ and $V_0^{(2)}$ there analyzed,
although it is stressed that the second one is a more reliable indication of the behavior of the heat propagation. 
In the integer quantum Hall effect, there is no fractionalization of the charge and the same electrons that transport
the charge current also transport  the energy along the edge. It is, thus, rather natural that the local voltage, which senses the electronic
propagation, is correlated to the local temperature, which senses the energy propagation.

\subsection{Quantum spin Hall systems}
\label{sec:qsh}
We now turn to analyze the case of a topological insulator, where there are a couple of edge states with electrons moving with
different chiralities and spin polarizations. In this case, we have
\[ \alpha(\omega)= \left\{ \begin{array}{ll}
        \gamma_+(\omega) + \gamma_- (1+\overline{\Gamma}_{2,-}) & \mbox{if $x<x_1, \;\; x \geq x_2$};\\
\gamma_+(\omega)      
(1+\overline{\Gamma}_{2,+})+ \gamma_-(\omega)  & \mbox{if $x_1 \leq x < x_2 $}.\end{array} \right. \] 
with
\begin{equation}
\gamma_{\sigma}(\omega)= \frac{\pi \overline{\Gamma}_{1,\sigma}}{4 \Lambda |\Delta_{\sigma}(\omega)|^2 \sin^2(\frac{\omega L}{2 v_F}+\frac{\Phi}{2})}
\end{equation}
Thus, the sensed voltage and temperature will change along the edge, provided that the
drain and/or source reservoirs are spin-polarized, in which case $\overline{\Gamma}_{j, +}
\neq \overline{\Gamma}_{j,-}$. For reservoirs without a net spin polarization, each branch of the
edge contains an identical flow
of electrons thermalized with the source reservoir propagating clockwise and 
electrons thermalized with the drain propagating with the opposite chirality. The result is a uniform voltage and temperature along
the edge, while the voltage and temperature drop in relation to the voltages and 
temperature of the reservoirs takes place at the contacts.

\section{Summary and conclusions}
\label{sec:conclusions}

In this paper we presented a simple model of a macroscopic droplet of a two-dimensional electron gas in the integer quantum Hall regime driven by an external ac source which acts as a heater. We used non-equilibrium methods to show that the heat current flows downstream (as expected) and derived expressions for the local voltage and temperature along the chiral edge of free fermions. Our results verify the arguments given by Granger and coworkers \cite{granger} as an interpretation of their experiments. We found, also in agreement with these experiments,  that the electrons along the
edge thermalize with the closest  ``upstream''  reservoir.

Several comments are in order. In this model the edge electrons are treated as free fermions while a more realistic description of the edge states of the integer quantum Hall systems should include the effects of electron-electron interaction. Interaction effects  should not change the results in an essential way since  the edges are chiral and interactions only affect the forward propagation of the fermions and no backscattering processes can occur at the contacts. Thus in the thermodynamic limit the temperature along the edges should remain constant (with only oscillations due to finite-size effects such as the ones we find here.) However it is possible that Coulomb interactions may lead to  oscillations  along the edge that may not vanish in the thermodynamic limit (as in the case of a quantum wire studied by Schulz\cite{Schulz-1993}.) 

It would be very interesting to have experiments of this type done in graphene devices since in graphene the integer quantum Hall effect is seen even at room temperature. Thus, graphene could prove to be an ideal system for testing these type of questions. Experiments of this type could also be used to test the basic physics behind the quantum spin Hall effect. Indeed, if the reservoirs are not polarized, no chiral heat current would be observable. In contrast, chiral spin currents should be detectable if the reservoirs are polarized (magnetized).


\begin{acknowledgments}
We thank G. Lozano and C. Na\'on for discussions. We acknowledge support from CONICET, ANCyT, UBACYT (Argentina), and the
J. S. Guggenheim Memorial Foundation (LA). LA thanks the ICMT of the University of Illinois for hospitality, and EF thanks Programa Ra{\'\i}ces 
(MINCYT, Argentina)  for support and the Department of Physics, FCEyN UBA (Argentina)  for hospitality. This work was supported in part by the 
National Science Foundation, under grants DMR 0758462 and DMR-1064319 (EF). 
\end{acknowledgments}

\appendix

\section{Retarded unperturbed Green functions} \label{apa}
In this section we evaluate the equilibrium Green functions $G^0_{x,x^{\prime}, \sigma}(\omega) $, corresponding to the edge in
contact to the source and drain reservoirs but free from the effect of the ac-driving voltage. This Green function can be evaluated 
from the solution of the following Dyson equation
\begin{equation}\label{set}
G^0_{x,x^{\prime},\sigma}(\omega)=g_{x,x^{\prime},\sigma}(\omega)+\sum_{j=1}^2 G^0_{x,x_j,\sigma}(\omega)\Sigma_{j,\sigma}(\omega)
g_{x_j,x^{\prime},\sigma}(\omega),
\end{equation}
with
\begin{equation}
g_{x,x^{\prime},\sigma}(\omega)=\frac{1}{\cal N} \sum_p \frac{1}{\omega - \sigma v_F (p-\frac{\Phi}{L})},
\end{equation}
the Green function of the free edge of chiral electrons, where $g_{x,x^{\prime},+}(\omega)=g_{x^{\prime},x,-}(\omega)$, 
with $p=2 n \pi /L, \;\; -K \leq n \leq K$, and ${\cal N}=2K+1$. In the limit $ K \rightarrow \infty$, this function reads
\begin{eqnarray}
g_{x,x^{\prime},+}(\omega)& = & \frac{ \pi e^{i\frac{\omega}{v_F} (x-x^{\prime})}}{2 \Lambda\sin(\frac{\omega L}{2 v_F}+ \frac{\Phi}{2})}
\{\Theta(x-x^{\prime}) e^{-i(\frac{\omega L}{2 v_F} +\frac{\Phi}{2} )} \nonumber \\
& + & \Theta(x^{\prime}-x) e^{i(\frac{\omega L}{2 v_F}+\frac{\Phi}{2} )} \},
\end{eqnarray}
 where $\Lambda$ is a high-energy cutoff.\cite{greenf} 
Assuming $x_1 < x_2$, the solution of Eq.\eqref{set} is
\begin{widetext}
\[ G^0_{x,x_1,+} (\omega)= \left\{ \begin{array}{ll}
         \frac{\pi e^{i\frac{ \omega}{v_F} (x-x_1)} e^{-i(\frac{\omega L}{2 v_F} +\frac{\Phi}{2} )}}
{2 \Lambda\sin(\frac{\omega L}{2 v_F}+ \frac{\Phi}{2})\Delta_{+}(\omega)} & \mbox{if $x<x_1, \;\; x \geq x_2$};\\
 \frac{\pi e^{i\frac{\omega}{v_F} (x-x_1)} e^{ i(\frac{\omega L}{2 v_F} +\frac{\Phi}{2} ) }}
{2 \Lambda\sin(\frac{\omega L}{2 v_F}+ \frac{\Phi}{2})\Delta_{+}(\omega)}         
(1+\overline{\Gamma}_{2,+}) & \mbox{if $x_1 \leq x < x_2 $}.\end{array} \right. \] 
and
\[ G^0_{x,x_2,+} (\omega)= \left\{ \begin{array}{ll}
         \frac{\pi e^{i\frac{\sigma \omega}{v_F} (x-x_2)} e^{-i(\frac{\omega L}{2 v_F} +\frac{\Phi}{2})} }
{2 \Lambda\sin(\frac{\omega L}{2 v_F}+ \frac{\Phi}{2})\Delta_{+}(\omega)} (1+\overline{\Gamma}_{1,+}) & \mbox{if $x<x_1, \;\; x \geq x_2$};\\
 \frac{\pi e^{i\frac{\omega}{v_F} (x-x_2)} e^{i(\frac{\omega L}{2 v_F} +\frac{\Phi}{2}) } }
{2 \Lambda\sin(\frac{\omega L}{2 v_F}+ \frac{\Phi}{2})\Delta_{+}(\omega)}         
 & \mbox{if $x_1 \leq x < x_2 $}.\end{array} \right. \]
\end{widetext}
where $G^0_{x,x^{\prime},+} (\omega)=G^0_{x^{\prime},x,-} (\omega)$, 
with
$\overline{\Gamma}_{j,\sigma}=\pi \Gamma_{j,\sigma}/\Lambda$ and
\begin{eqnarray}
\Delta_{\sigma}(\omega)&=&[1-\Sigma_{1,\sigma}g_{x_1,x_1,\sigma}(\omega)][1-\Sigma_{2,\sigma}g_{x_2,x_2,\sigma}(\omega)]
\nonumber \\
& & -
\Sigma_{1,\sigma}g_{x_1,x_2,\sigma}(\omega)\Sigma_{2,\sigma}g_{x_2,x_1,\sigma}(\omega).
\end{eqnarray}

\end{document}